\documentclass[twocolumn,floatfix,showpacs,pra,aps,letterpaper]{revtex4}

\usepackage{amsmath}
\usepackage{amssymb}
\usepackage{graphicx}
\usepackage{epsfig}
\usepackage{color}
\usepackage{multirow}

\newcommand{\CNOT}{\textsc{cnot}}
\newcommand{\SWAP}{\textsc{swap}}
\newcommand{\CPHASE}{\textsc{cphase}}

\begin{document}

\title{Accuracy threshold for concatenated error detection in one dimension}

\author{Ashley M. Stephens$^{}\footnote[1]{Electronic address: a.stephens@physics.unimelb.edu.au}$ and Zachary W. E. Evans}

\affiliation{
Centre for Quantum Computer Technology, School of Physics\\
University of Melbourne, Victoria 3010, Australia.}
\date{\today}

\begin{abstract}
Estimates of the quantum accuracy threshold often tacitly assume that it is possible to interact arbitrary pairs of qubits in a quantum computer with a failure rate that is independent of the distance between them. None of the many physical systems that are candidates for quantum computing possess this property. Here we study the performance of a concatenated error-detection code in a system that permits only nearest-neighbor interactions in one dimension. We make use of a new message-passing scheme that maximizes the number of errors that can be reliably corrected by the code. Our numerical results indicate that arbitrarily accurate universal quantum computation is possible if the probability of failure of each elementary physical operation is below approximately $10^{-5}$. This threshold is three orders of magnitude lower than the highest known.
\end{abstract}

\pacs{03.67.Lx, 03.67.Pp}
\maketitle

\section{Introduction}

For a quantum computer to reliably outperform a classical computer, it must be robust against the effects of decoherence and imprecise quantum control. There are many ideas on how to achieve such fault tolerance, including topological quantum computing \cite{Kitaev1, Nayak1, Raussendorf2}, surface codes \cite{Kitaev1, Bravyi1, Dennis1}, color codes \cite{Bombin1}, self-correcting codes \cite{Bacon1}, and concatenated codes \cite{Shor2, Steane1}. The threshold theorem indicates that, under certain conditions, scalable quantum computing is possible in principle \cite{Aharonov1, Aliferis1}. The theorem asserts that if the probability of failure of each elementary physical operation in a quantum computer is below some threshold then arbitrarily accurate quantum computation can be performed efficiently given sufficient time and qubits. The actual value of the threshold for a given error-correction code depends on a number of parameters that describe the quantum computer and the noise that affects it \cite{Gottesman1}.

In most estimates of the accuracy threshold, the assumption is made that it is possible to interact arbitrary pairs of qubits in a quantum computer with a failure rate that is independent of the distance between them. This property is desirable, since it allows higher failure rates to be tolerated, but it is unrealistic in the limit of many qubits. In all of the physical systems that are candidates for quantum computing the range of controllable interactions is constrained such that at least some pairs of qubits will need to be transported before they can undergo logic gates such as $\CNOT$ and $\CPHASE$. Moreover, we expect that in many systems, especially trapped-ion, superconducting, and solid-state systems, hardware limitations will require that the qubit array is two- or one-dimensional \cite{Kielpinski1, Taylor1, Hollenberg1, Fowler1}.

The combination of local interactions and low coordination number is not a problem in principle as it is known that the threshold theorem still applies \cite{Aharonov1, Gottesman4}. The value of the threshold has been quantified for systems that permit only nearest-neighbor interactions in two dimensions \cite{Raussendorf1, Svore1, Roychowdhury1} and in various quasi one-dimensional settings \cite{Szkopek1, Stephens1}, and the threshold is also known for superconducting \cite{Fowler1} and ion-trap \cite{Metodi1} architectures with similar properties. However, there are a large number of systems under development that permit only nearest-neighbor interactions in one dimension \cite{Kane1, Loss1, Vrijen1, Tian1, Hollenberg2, Feng1, Pachos2, Vandersypen1, Solinas1, Jefferson1, Petrosyan1, Vyurkov1, Kamenetskii1}, for which the threshold is unknown. The threshold in one dimension is expected to be significantly lower than in all other cases.

Here we find the accuracy threshold for a system that permits only nearest-neighbor interactions in one dimension. Where required, qubits are transported via nearest-neighbor $\SWAP$ gates. To minimize this overhead we use a small error-detection code. To correct errors we make use of a new message-passing method that uses classical information gathered during error detection to maximize the number of errors that can be reliably corrected \cite{Evans2}. Our numerical results indicate that arbitrarily accurate quantum computation is possible if the probability of failure of each elementary physical operation is below approximately $10^{-5}$.

\section{[[4,1,2]] subsystem code}

The code that we have chosen to use is the [[4,1,2]] subsystem code, a stabilizer CSS quantum code that encodes one logical qubit into four physical qubits \cite{Bacon1, Bacon2}. Its stabilizer, $\mathcal{S}$, is
\begin{equation}
\begin{aligned}
X_1X_2X_3X_4, Z_1Z_2Z_3Z_4,
\end{aligned}
\label{stabilizers:bs}
\end{equation}
where $X_i$ and $Z_i$ represent the Pauli operators $\sigma_X$ and $\sigma_Z$ applied to the $i^{th}$ qubit respectively. Identity operators and tensor products between operators are omitted. Although there are two degrees of freedom in which to store encoded information, we choose to ignore one of these encoded qubits. Then, elements in the non-Abelian group, $\mathcal{T}$, generated by the operators
\begin{equation}
\begin{aligned}
X_1X_2, X_3X_4, Z_1Z_3, Z_2Z_4,
\end{aligned}
\label{gauge}
\end{equation}
act trivially on the sole protected qubit. The encoded $Z$ and $X$ operators for this qubit are
\begin{equation}
X_1X_3, Z_1Z_2.
\end{equation}
Because the elements in $\mathcal{S}$ are products of elements in $\mathcal{T}$, and since all elements in $\mathcal{T}$ commute with all elements in $\mathcal{S}$ and the encoded operators, to determine the eigenvalues of the elements of $\mathcal{S}$ it is sufficient to measure the eigenvalues of the elements in $\mathcal{T}$. This may change the state of the system but it will not affect the state of the protected qubit. This property allows us to use only two ancilla qubits to simultaneously perform operator measurements of the operators in $\mathcal{T}$ \cite{Aliferis3}. We decode the syndrome by taking the parity of the two measurement outcomes in each basis. This is used to infer the presence or absence of errors in each basis---if the parity is even there is no error, if the parity is odd there is an error. For the [[4,1,2]] subsystem code, $\CNOT$ is a valid transversal gate and $H$ is a valid transversal gate up to a permutation of qubits.

\section{Using concatenated error detection to correct errors}

In our scheme the [[4,1,2]] subsystem code is concatenated such that physical qubits form encoded qubits which in turn form higher-level encoded qubits and so on. However, while concatenating the code $l$ times results in a final code that has distance $2^{l+1}$, if each level of error detection operates independently of every other level, then the code cannot reliably correct even a single error---a single physical failure may cause an encoded error at any level. To do more than simply detect errors classical messages must be passed from each level to the level above. These messages serve to indicate the location of potential errors, thereby removing the ambiguity in the cause of any odd parity syndrome that is observed, which allows errors that are detected to be located and corrected. 

The message-passing method applied in this paper is similar to those of Refs.~\cite{Knill1, Evans1, Aliferis5} but, unlike those, it will only fail if the number of concurrent errors is greater than or equal to half of the distance of the final concatenated code---in the case of the [[4,1,2]] subsystem code, the leading-order exponent of the probability of failure scales with the number of levels of concatenation as $1 $(physical)$,1,2,4,8,16,32$, and so on. This is also a property of the noisy-channel method of Ref.~\cite{Poulin1}. Although we will only describe the message-passing method in the case where it is applied to the [[4,1,2]] subsystem code, it can be applied to codes with greater distance. A detailed description of the method is contained in Ref.~\cite{Evans2}.

Messages in the method consist solely of classical information which indicates our confidence that each location in the circuit has not failed given what is known about it. We call these messages flags. The probability that any given flag represents an actual error is described by the weight, $w$, carried by the flag such that $\textrm{Pr}(\textrm{error}) = \mathcal{O}(p^w)$, where $p$ is the probability of failure of an elementary physical location in the quantum computer. Since we assume that errors at each location are independent of all other locations, the probability of a set of flags representing actual errors is $\mathcal{O}(p^{\sum{w}})$, where the sum is over the entire set. Flags are raised at every elementary physical and encoded location. All flags at the physical level are given weight equal to one by definition and at all other levels weights are determined during error detection at the level below. Flags in the $X$ and $Z$ bases are separate.

As flags are raised they are classically propagated through the error-detection circuit to determine the effect that an error at the location at which the flag originated would have on the data and ancilla qubits at the point of syndrome extraction. Note that $\CNOT$ copies $X$ errors from the control qubit to the target qubit and $Z$ errors in the opposite direction, and that we define the point of syndrome extraction to be immediately after the ancilla qubits are measured. Since there are significantly fewer effects than there are locations at which flags are raised, flags are binned by effect. Each bin is assigned a single weight which is only updated if a flag is raised with a weight lower than its existing value. 

\begin{figure*}
\begin{center}
\resizebox{100mm}{!}{\includegraphics[angle=270]{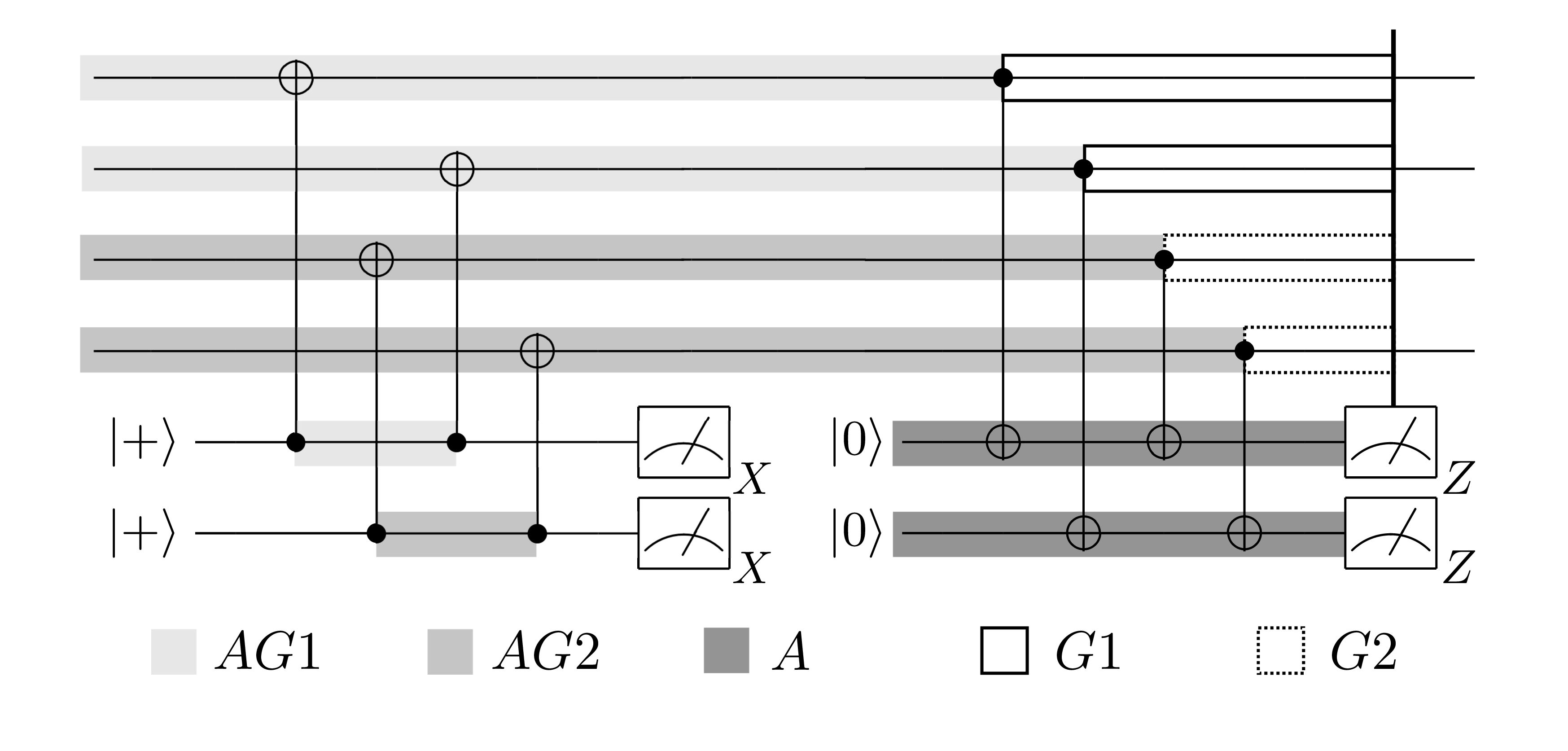}}
\end{center}
\vspace*{-20pt}
\caption{Flags are binned according to the effect that an error at the location at which the flag originated would have on the data and ancilla qubits at the point of syndrome extraction. For simplicity, only $X$ errors are considered. An $X$ error in the bin $AG1$ would change the parity of the ancilla while remaining on one of the first pair of data qubits, an $X$ error in $AG2$ would change the parity of the ancilla while remaining on one of the second pair of data qubits, and an $X$ error in $A$ would change the parity of the ancilla.  Errors in $G1$ and $G2$ cannot affect parity of the ancilla, but the weights of $G1$ and $G2$ are recorded and used to update the weights of $AG1$ and $AG2$ for the next error-correction cycle. Binning of flags corresponding to $Z$ errors is undertaken similarly.}
\label{figure:bi}
\end{figure*}

Figure \ref{figure:bi} illustrates the binning system where, for simplicity, non-local circuits are used and only $X$ errors are considered. Note that the $X$-error syndrome is obtained using the second half of the circuit in Fig.~\ref{figure:bi}, so we are interested in the effect of $X$ errors at the second set of measurements. $X$ errors do not affect the $Z$-error syndrome, but since they can occur during the circuit to obtain the $Z$-error syndrome this part of the circuit must be considered. Because of the degeneracy of the code, we need only consider two pairs of data qubits rather that four individual data qubits. For example, since $X_1X_2$ acts trivially on the encoded qubit, both $X_1$ and $X_2$ give the same syndrome and can be corrected by the operator $X_1$.

Since the failure of a single $\CNOT$ or $\SWAP$ gate can introduce a pair of correlated errors, these gates require special treatment. The weights of the two flags that are raised immediately following each two-qubit gate are used to update the bin which corresponds to the two-qubit correlated error. Note that both of the bins corresponding to single errors following the two-qubit gate are also updated. The weight of the correlated error is taken to be the maximum of the two single-qubit weights, as opposed to the sum, which would describe the probability of the pair of errors occurring without correlation.

Once the syndrome is measured its most likely cause is identified by finding the bin that has the lowest weight while still being consistent with, or matching, the syndrome. In the case of the [[4,1,2]] subsystem code, in each error-correction cycle we consider only three bins, $AG1$, $AG2$, and $A$. The most likely cause of an odd syndrome is a single error occurring in the bin with the lowest weight, as each of these bins corresponds to a change in the parity of the ancilla. The most likely cause of an even syndrome is always no error at all. If the match implies an error on a data qubit then the appropriate correction is applied.

Although we correct for the most likely error it is possible that the true error is the complement of the correction---that is, the true error and the correction that we apply combine to form an encoded operator. To determine the weight of the flag that is raised in the error-correction circuit at the level above, what we will call the encoded weight, we calculate the difference between the weight of the match on which we act and the weight of its complement. For example, if we have an odd syndrome and the $AG1$ is the match, then $AG2$ is the complement match and the encoded weight is the difference between the weights of the these two matches, $AG2-AG1$. 

There is the possibility that a correction will result in a state that is outside the code space so that it is neither correct nor affected by an encoded operator. Just as the complement match is used to determine the probability of an encoded error, the conditional probability of single errors on the pairs of data qubits can be updated by considering other matches. At the end of each error-correction cycle the weights of $AG1$ and $AG2$ are updated to the minimum of the weight of any previous locations that have not yet had an opportunity to affect the ancilla, given by $G1$ and $G2$, and a weight obtained during the preceding circuit, calculated in an analogous way to the encoded weight based on the syndrome that is observed.

Table I shows a list of all possible flag matches along with the various flag-weight updates that would result from each of them being acted on, where we have retained the notation of Fig.~\ref{figure:bi}. By the careful consideration of every possible cause of every syndrome in this way we always have accurate weights for every qubit at every level. We can always, therefore, apply corrections based on the most likely set of errors.
\begin{table*}
\begin{tabular}{c|c||c|c|c|c}
  synd. & match & corr. & $C_{1}$ & $C_{2}$ & $C_{e}$ \\
  \hline
  \multirow{4}{*}{+1}
   & None & none & $AG1+A$ & $AG2+A$ & $AG1+AG2$\\
   & $AG1+A$ & \multicolumn{4}{c}{n/a}\\
   & $AG2+A$ & \multicolumn{4}{c}{n/a}\\
   & $AG1+AG2$ & \multicolumn{4}{c}{n/a}\\
   \hline
  \multirow{4}{*}{-1}
  & $AG1$ & $X_1/Z_1$ & $A-AG1$ & $AG2+A$ & $AG2-AG1$\\
  & $AG2$ & $X_3/Z_2$ & $AG1+A$ & $A-AG2$ & $AG1-AG2$\\
  & $A$ & none & $AG1-A$ & $AG2-A$ & $AG1+AG2$\\
  & $AG1+AG2+A$ & \multicolumn{4}{c}{n/a}
  \label{tab: flag matches}
\end{tabular}
\caption{Table of all flag matches for each of the two possible syndromes \cite{Evans2}. The first column lists the syndrome. The second column lists the possible flag matches for each syndrome. The correction and flag update rules, shown in the last four columns, depend on which is the lowest-weight flag match. $C_{1}$ and $C_{2}$ are weights for $G1$ and $G2$ at the current level of error correction, which become $AG1$ and $AG2$ respectively for the next error-correction cycle. $C_e$ is the encoded weight for the next-highest level of error correction.}
\end{table*}

\section{Accuracy threshold in one dimension}

To estimate the threshold for universal quantum computation in one dimension we construct circuits for error detection and for the encoded operations $\CNOT$, $\SWAP$, Hadamard, state preparation, and measurement.  These operations, which we will refer to as CSS operations, are sufficient to concatenate error detection and to perform state distillation following the ideas presented in Refs.~\cite{Bravyi3,Aliferis4}. State distillation involves preparing the ancillary state $\vert{\textrm{+}i}\rangle=\vert0\rangle+i\vert1\rangle$ to enable the logical phase gate, $S$, and preparing the ancillary state $\vert\textrm{Toffoli}\rangle=\vert000\rangle+\vert010\rangle+\vert100\rangle+\vert111\rangle$ to enable the logical Toffoli gate. These gates together with the CSS operations complete a universal set for quantum computation. Accurate states can be distilled from many noisy states provided the noisy states can be made with a failure rate lower than some distillation threshold, which is typically above one percent.  We determine the threshold for CSS operations by numerically simulating the circuit for an error-corrected logical $\CNOT$ under a stochastic error model. The $\CNOT$ is chosen because it has the highest failure rate of the CSS operations. As the threshold we find is well below the distillation threshold it is the threshold for universal computation under our scheme.

\begin{figure}
\begin{center}
\resizebox{63mm}{!}{\includegraphics{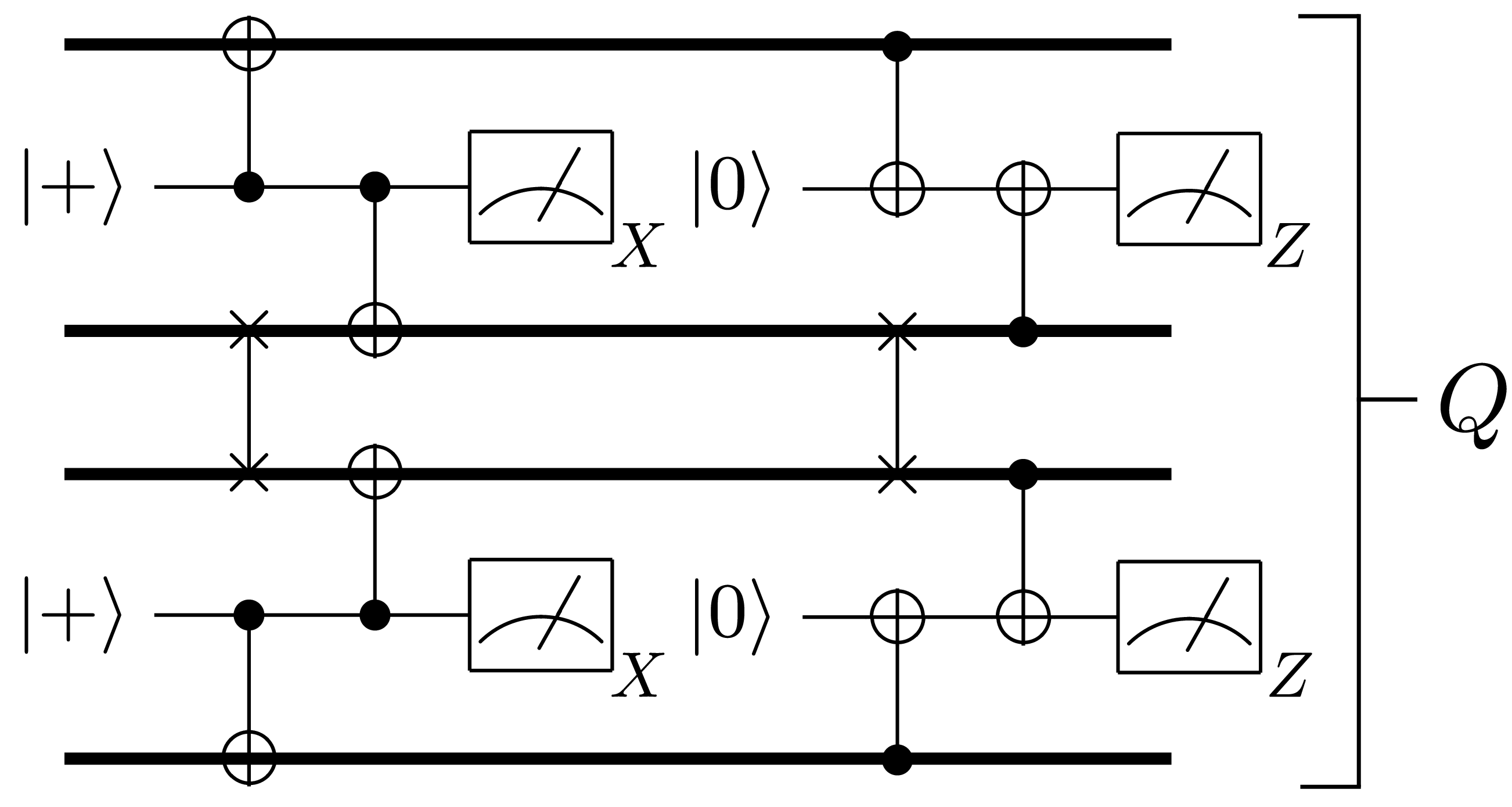}}
\end{center}
\vspace*{-10pt}
\caption{Syndrome-extraction circuit for the [[4,1,2]] subsystem code on a linear nearest-neighbor array. $Q$ denotes the configuration of the encoded qubit after syndrome extraction, bold lines indicate data qubits, and plain lines indicate ancilla qubits.}
\label{figure:se}
\end{figure}

Figure \ref{figure:se} shows a circuit for syndrome extraction for the [[4,1,2]] subsystem code where only nearest-neighbor interactions on a linear array are permitted. Note that the encoded Hadamard can be achieved by transversal application of the Hadamard gate in addition to removing a single $\SWAP$ gate from this circuit to permute the qubits. This syndrome-extraction circuit has the same depth as the non-local circuit in Fig.~\ref{figure:bi} and the two circuits differ only by a rearrangement of qubits and the addition of two $\SWAP$ gates. This difference is significant, however, since each of the additional $\SWAP$ gates involves two data qubits. New pairs of correlated errors are introduced with probability $\mathcal{O}(p)$, where $p$ is the probability of failure of an elementary physical location. Without the $\SWAP$ gates these particular pairs of errors occur with probability $\mathcal{O}(p^2)$. Since these errors include the encoded operators, this means that an undetected encoded error occurs with probability $\mathcal{O}(p)$.

\begin{figure}
\begin{center}
\resizebox{57mm}{!}{\includegraphics{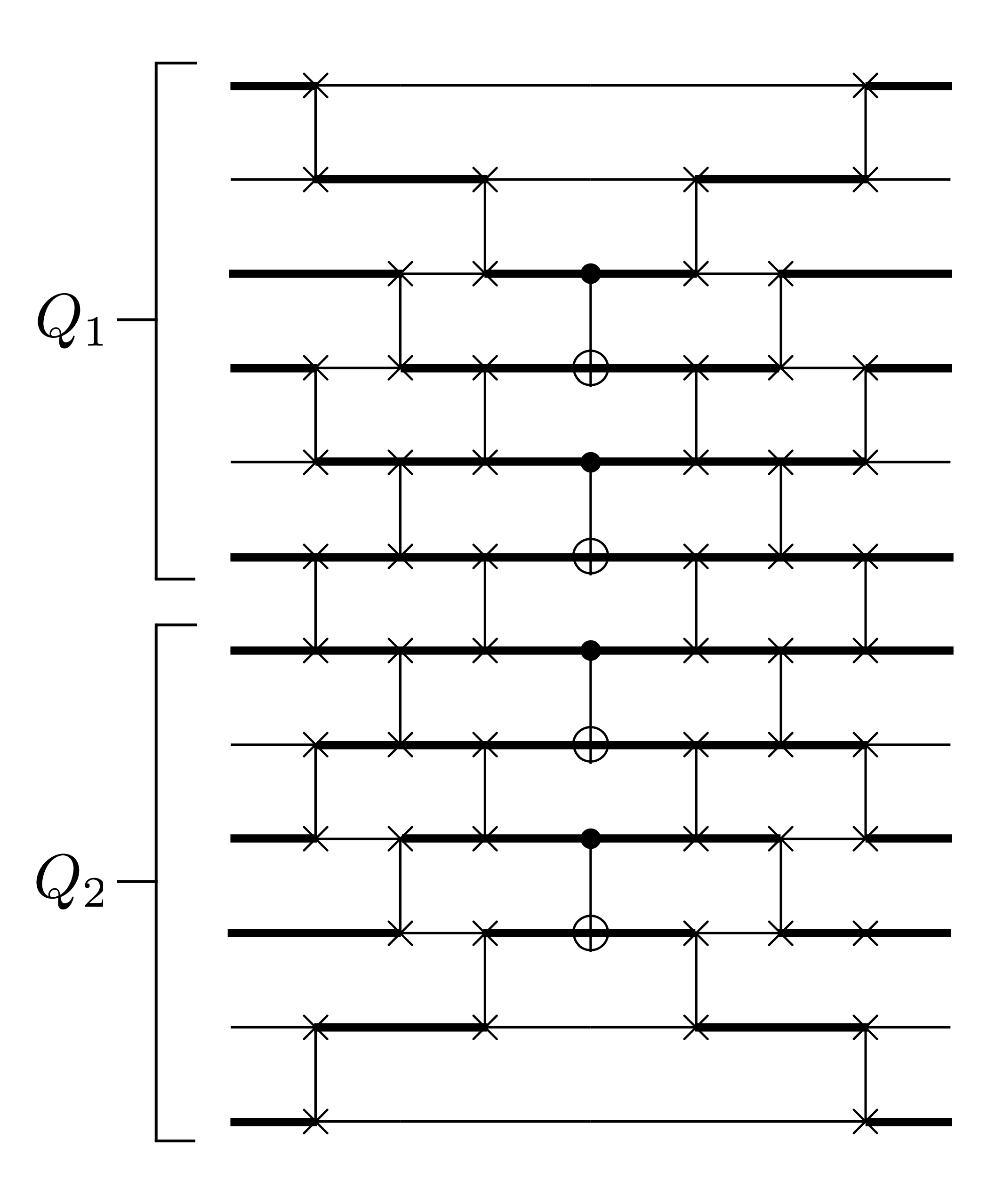}}
\end{center}
\vspace*{-20pt}
\caption{Encoded $\CNOT$ circuit for the [[4,1,2]] subsystem code on a linear nearest-neighbor array, where notation is consistent with that of Fig.~2. Replacing all $\CNOT$ gates with $\SWAP$ gates gives the encoded $\SWAP$ circuit.}
\label{figure:cn}
\end{figure}

Figure \ref{figure:cn} shows the encoded $\CNOT$ circuit. A pair of correlated errors caused by the failure of one of the $\SWAP$ gates prior to the transversal $\CNOT$ results in a pair of errors on the encoded control qubit and a pair of errors on the encoded target qubit. This is because the transversal $\CNOT$ copies errors. This means that a pair of correlated, undetected encoded errors occurs with probability $\mathcal{O}(p)$. The encoded $\SWAP$ gate, in which the transversal $\CNOT$ is replaced by a transversal $\SWAP$, does not possess this property. Two gates must fail during the encoded $\SWAP$ for both encoded qubits to fail undetected. 

Here we make an important observation: to leading order, the relative probabilities of undetected failure of first-level encoded locations in the one-dimensional case mimic the relative probabilities of failure of physical locations in the non-local case. Specifically, all single-qubit locations fail with probability $\mathcal{O}(p)$, a pair of correlated errors after a $\CNOT$ occurs with probability $\mathcal{O}(p)$, and a pair of correlated errors after a $\SWAP$ occurs with probability $\mathcal{O}(p^2)$. A pair of correlated errors can occur after a $\SWAP$ gate because of a single fault, but they will always be detected. A corollary of this is that after the first level of concatenation our linear scheme will mimic a non-local scheme. In the non-local case we expect to succeed whenever the number of concurrent errors is less than half of the distance of the final concatenated code. This implies that in the one-dimensional case the first-order exponent of the probability of failure should scale with the number of levels of concatenation as $1 $(physical)$, 1,1,2,4,8,16$, and so on.

Note that Figs.~\ref{figure:se} and \ref{figure:cn} show circuits constructed from physical gates. To generate circuits for error correction and encoded operations at higher encoded levels we replace all physical gates with encoded gates. Like in other concatenated schemes, after every encoded operation we perform error correction on the encoded qubits involved in that operation \cite{Aliferis1}. All circuits are designed so that ancilla qubits are always available to perform error correction using the circuit in Fig.~\ref{figure:se}.

To attempt to verify that our scheme performs as expected and to determine the threshold, we simulate a logical $\CNOT$ under a stochastic error model. The simulated circuit consists of a $\CNOT$ extended rectangle \cite{Aliferis1}, the failure rate of which is meant to approximate the failure rate of a $\CNOT$ in some algorithm or at a higher level of concatenation. Because the circuit only contains gates from the Clifford group, we need only simulate the propagation of errors that occur during the circuit \cite{Steane3} rather than store the complete state of the quantum computer. We have written our own simulator for this purpose. We simply assume the state begins in an arbitrary valid codeword state and is stochastically perturbed by errors during the circuit. The circuit is defined to succeed if measurement of the data qubits at the end of the circuit in either the $X$ or $Z$ basis would give the correct outcome. Equivalently, a circuit is defined to have succeeded if an errorless error-correction cycle applied to its output state would produce the correct state.

We generate data in two different ways. Where $n$ is the number of levels of error correction, for $n\leq3$, instead of applying an error at each location with some probability, we simulate the full concatenated circuit with exactly $i$ errors placed randomly. For all single-qubit locations (state preparation, memory, measurement) the error is a randomly selected Pauli error and for all two-qubit locations ($\CNOT$, $\SWAP$) the error is a randomly selected two-qubit Pauli error. This is repeated many times to generate the probability that the circuit fails given $i$ errors, $r_i$. These conditional probabilities can be combined to give the failure rate of the circuit as a function of $p$,
\begin{equation}
\sum^N_{i=0}{r_i {N\choose i}p^i(1-p)^{N-i}},
\label{eq: expansion}
\end{equation}
where $N$ is the number of locations in the entire circuit, 172, 11992, and 864496 for $n=1,2,3$ respectively. The series is truncated after $i=6$, 10, and 21 for $n=1,2,3$ respectively. These numbers are chosen so that the approximation is valid in the region of the threshold.

For $n=4$, too much time is required to generate statistically significant results in this way. Instead, after each elementary physical location in the circuit we apply an error with probability $p$, where errors at all locations are independent. We study the failure rate of the circuit as we vary $p$ between $8\times10^{-6}$ and $2\times10^{-5}$, again, simulating the full concatenated circuit. The time required to generate statistically significant results for $n>4$ is prohibitive using our methods. In our simulator we use the SIMD-oriented Mersenne Twister pseudorandom number generator \cite{Saito1}. The message-passing method is simulated along side the error-correction circuit so that it operates in the same way as it would in a real quantum computer. 

\begin{figure*}[t!]
\begin{center}
\resizebox{105mm}{!}{\includegraphics{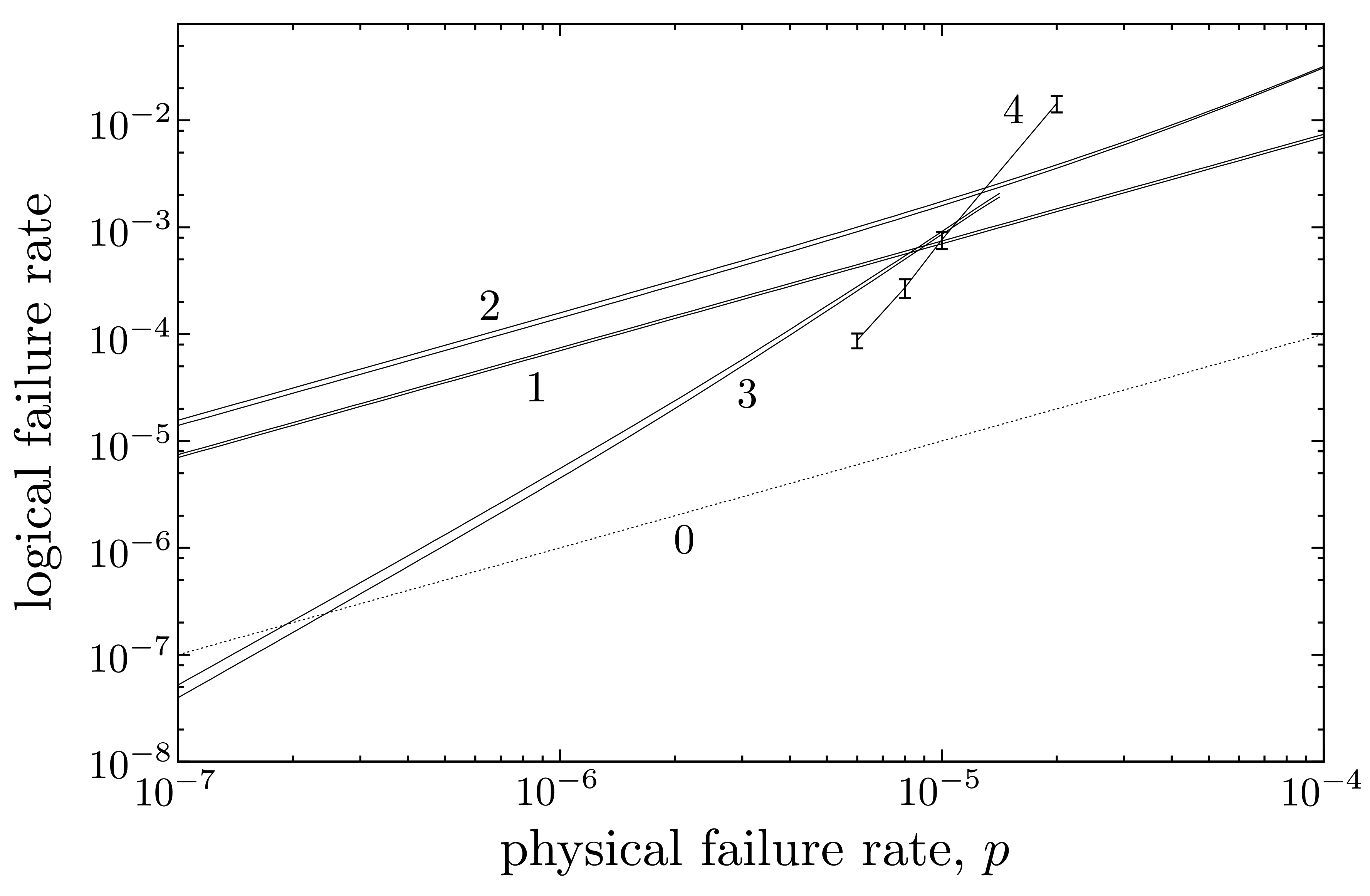}}
\end{center}
\vspace*{-15pt}
\caption{Logical failure rate as a function of physical failure rate for up to four levels of error correction (lines labeled 1$-$4). No error correction (dashed line labeled $0$) is shown for reference. Lines for $n=1,2,3$ were calculated using the expansion in Eq.~4. To show the statistical error in the parameters $r_i$ there are two lines for $n=1,2,3$. The top line in each pair is $2\sigma$ above the mean and the bottom is $2\sigma$ below the mean. Data for $n=4$ was obtained directly. The line for $n=4$ joins the means of the three data points, where the error bars indicate $\pm 2\sigma$ statistical error. The threshold is the value of $p$ at which the failure rate of the $n^{th}$ level of error correction intersects the physical failure rate in the limit $n \rightarrow \infty$. Assuming that the lines corresponding to higher-level failure rates continue the trend observed, the threshold will be close to the crossing of lines for $n=3$ and $n=4$ at approximately $10^{-5}$.}
\label{figure:re}
\end{figure*}

Our results are summarized in Fig.~\ref{figure:re}. The gradients of the lines in Fig.~\ref{figure:re}, which are related to the minimum number of errors that cause the code to fail, are as expected. Since the first two levels of error correction are unable to reliably correct errors, the failure rate for any less than three levels of error correction is always greater than $p$. Note that the lines for $n=3$ are truncated at $p>10^{-5}$. This is because for $p>10^{-5}$ the approximation that $r_i=0$ for $i>21$ breaks down as failure due to more than 21 errors becomes common.

Our results suggest that the threshold is approximately equal to $10^{-5}$. For elementary physical failure rates below this threshold, we expect that arbitrarily accurate CSS operations can be performed efficiently given sufficient time and qubits. This threshold is well below the threshold for distillation of ancilla states that enable universal computation \cite{Bravyi3,Aliferis4}. A failure rate much lower than $10^{-5}$ would, however, be required to achieve a sufficiently low encoded failure rate using a practical amount of resources. For example, we estimate that to achieve an encoded failure rate of $10^{-15}$ using $\mathcal{O}(10^3)$ physical qubits per encoded qubit, a physical failure rate of approximately $10^{-8}$ is necessary. Furthermore, if there is a non-zero probability of defective qubits in the linear array then the threshold will cease to exist. In contrast, a constant density of defective qubits in a two-dimensional array can be tolerated and will merely lower the threshold.

\section{Conclusions and further work}

That the accuracy threshold for universal quantum computation in a system that permits only nearest-neighbor interactions in one dimension may be $10^{-5}$ or higher is somewhat surprising. In two dimensions the highest proven threshold for concatenated error correction is $2\times10^{-5}$ \cite{Roychowdhury1}. There is strong evidence that the threshold in two dimensions can be as high as $7\times10^{-3}$ \cite{Raussendorf1} but this relies on techniques that are not expected to be useful in only one dimension. In quasi one-dimensional settings the highest proven threshold is $2\times10^{-6}$ \cite{Stephens1}. Since the threshold presented in this paper is based on numerical simulations, it would be useful to obtain a rigorous bound on its value. Due to the unconventional error-correction method that we use, it is unclear if this can be done using the established level-reduction procedure of Ref.~\cite{Aliferis1}.  

By adding $\SWAP$ gates where necessary, any quantum algorithm can be implemented on a linear nearest-neighbor array with an overhead that is, at most, polynomial in the number of qubits. This has been done explicitly for Shor's algorithm \cite{Shor1} in Ref.~\cite{Fowler2}. Our result strengthens the notion that one-dimensional architectures are viable candidates for quantum computing, although how to achieve defect tolerance remains an open question. It will be interesting to see if a higher threshold can be achieved by adapting the postselection scheme of Ref.~\cite{Knill1} to a system that permits only nearest-neighbor interactions in one dimension. In this scheme ancilla states are protected by an error-detection code and postselected. We expect that the new methods of message passing presented in Ref.~\cite{Evans2} will help in improving the efficiency of such a scheme, as the weights outputted from error correction may be useful in moderating the amount of postselection. These ideas are the subjects of further work. 

\section*{Acknowledgements} We thank Charles Hill and Magdalena Carrasco for their helpful suggestions. We acknowledge financial support from the Australian Research Council (ARC), the US National Security Agency (NSA), and the Army Research Office (ARO) under contract number W911NF-04-1-0290.

\bibliographystyle{unsrt}

\end{document}